\documentclass[a4paper]{jpconf}
\usepackage{graphicx}
\begin{document}
\title{Quantum Heisenberg antiferromagnetic chains with exchange
 and single--ion anisotropies}

\author{D Peters$^1$, I P McCulloch$^2$ and W Selke$^1$}
\address{$^1$ Institut f\"ur Theoretische Physik, RWTH Aachen University
  and JARA-SIM, 52056 Aachen, Germany}
\address{$^2$ Department of Physics, University of Queensland,
  Brisbane, QLD 4072, Australia}  

\ead{selke@physik.rwth-aachen.de}

\begin{abstract}
Using density matrix renormalization group calculations, ground state
properties of the spin--1 Heisenberg chain with
exchange and quadratic single--ion anisotropies in an external field
are studied, for special choices of the two kinds of
anisotropies. In particular, the phase diagram includes 
antiferromagnetic, spin--liquid (or spin--flop), (10), and
supersolid (or biconical) phases. Especially, new features of
the spin--liquid and supersolid phases are discussed. Properties
of the quantum chains are compared to those of corresponding classical
spin chains.
\end{abstract}

\section{Introduction}
Recently, low--dimensional quantum anisotropic Heisenberg
antiferromagnets in a field have been shown to display the
analog of the supersolid phase \cite{seng1,seng,lafl,picon,peters}. In
the language of classical magnetism, the phase is usually
denoted as the biconical phase \cite{fisher}, in which 
the order parameters of both the antiferromagnetic
and the spin--flop phase do not vanish. For instance, the
supersolid phase has been found to occur in the  ground state of
a spin--1 antiferromagnetic chain with exchange and single--ion
anisotropies, applying 
Monte Carlo simulations and perturbation theory \cite{seng}. The ratio
between the uniaxial exchange anisotropy $\Delta$ and the competing quadratic
single--ion anisotropy $D$ had been
fixed, $D/J= \Delta/2$ \cite{seng1,seng}. In our subsequent
study on the same model, using density matrix renormalization group (DMRG)
techniques, we provided evidence, for spin
chains of finite length, for having two types of supersolid as well as
two types of spin--liquid, known in classical magnetism 
as spin--flop, structures \cite{peters}. In this contribution, we shall
elaborate on our previous study. Moreover, we shall also investigate
another part of the ground state phase diagram by fixing
$\Delta$, $\Delta= 5$, and varying $D$, as had
been done before for a limited range 
of $D$ \cite{sakai}. Again, the
supersolid phase is stable. In addition, we 
identify commensurate and
incommensurate spin--liquid structures. We compare results
on the quantum chains to ones on corresponding classical spin
chains, discussing similarities and striking differences.

\section{Results}

In the following, we shall consider the spin--1 anisotropic
antiferromagnetic Heisenberg chain in a field described by the Hamiltonian

\begin{equation}
{\cal H} = \sum\limits_{i}
   (J (S_i^x S_{i+1}^x + S_i^y S_{i+1}^y
    + \Delta S_i^z S_{i+1}^z) + D (S_i^z)^2  - B S_i^z)
\end{equation}

\noindent
where $i$ denotes the lattice sites. For $\Delta >1$, there is an
uniaxial exchange anisotropy, along the direction of the
field, $B >0$, the $z$--axis. Depending on the
sign of $D$, the single-ion term leads to a competing
planar, $D > 0$, or to an enhancing uniaxial anisotropy. We shall
analyse ground
state properties using DMRG techniques \cite{white,scholl} for chains with
open boundary conditions and up to $L= 128$ sites. In addition, we
shall determine the ground states of corresponding infinite 
chains with classical spin vectors of length one \cite{holtwes,holt}. 

We first briefly deal with the case $D/J= \Delta/2$
\cite{seng,peters}. The ground state phase diagram in the
$(\Delta,B/J)$ plane has been found \cite{seng,peters} to comprise
antiferromagnetic (AF), spin--liquid (SL), supersolid
(SS), ferromagnetic (F), and (10), with a magnetization plateau
at half saturation, phases. At small values of $\Delta$ and
small fields, the Haldane phase is observed \cite{haldane}. In
comparison, the
corresponding classical spin chain shows a much broader 
biconical (BC) phase, being effectively replaced not only by the 
supersolid phase but also, largely, by the spin--liquid and (10)
phases. The classical (10) phase becomes stable only in the limit of an
Ising antiferromagnetic chain with a single--ion
term, the Blume--Capel model.  

Interesting information is given by the 
magnetization profiles, $m_i= <S^z_i>$, with brackets, $<...>$,
denoting quantum mechanical expectation values. In particular, we
observed distinct profiles in the SS phase in between the AF and
(10) and in between the AF and SL  phases, respectively. For odd
$L$, in the SS phase on approach to the (10) phase, the local magnetizations
$m_i$ at odd sites stay close to one, while at even sites they tend
roughly to zero. In
contrast, in the SS phase on approach to the SL phase, the
magnetizations on odd and even sites
tend to take on the same values \cite{peters}. In the SL phase we
found also (studying situations with $\Delta$ exceeding $\approx$ 2.5)
two distinct types of profiles, for finite chains, when
varying $M/L$, where $M$ is
the total magnetization \cite{peters}: For $M < L/2$, the
profiles exhibit a broad plateau in
the center of the chain, as expected for a classical spin--flop (SF)
structure, while pronounced modulations in $m_i$
occur at $M >L/2$, see Fig. 1. This may signal a change from commensurate (C)
to incommensurate (IC) structures \cite{peters}. The suggestion is
now confirmed and quantified by analysing the Fourier transform
of the profiles, especially at $\Delta= 3.5$. The
modulation in the IC region of the
SL phase is described nicely by the wavenumber (setting the lattice
spacing equal to one) $q= 2 \pi (1-m)$, $m= M/L$. Such
an IC modulation, with algebraic decay, is expected to
hold in the entire SL phase of the spin--1/2 anisotropic Heisenberg
chain in a field \cite{kimu}. In the classical variant, we find no
IC structures in the spin--flop phase, for finite and 
infinite chains.

\begin{figure}[h]
\includegraphics[width=16pc]{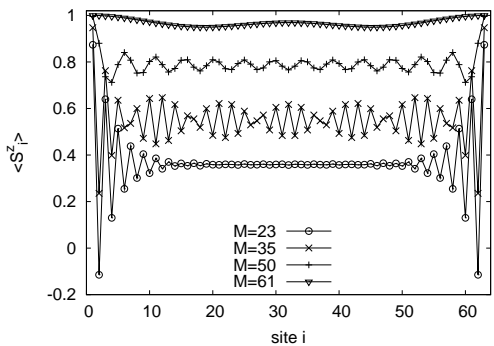}\hspace{2pc}%
\begin{minipage}[b]{16pc}\caption{\label{label}Magnetization profiles
in the spin--liquid phase for $\Delta =2D/J= 3.5$ at various total
magnetizations for a chain with $L= 63$ sites.}
\end{minipage}
\end{figure}

Let us now turn to the case of fixed exchange
anisotropy, $\Delta= 5.0$, varying the 
single--ion term, $D$. The ground state phase diagrams for
the spin--1 and the classical chains are depicted in Figs. 2 and
3, using DMRG calculations for chains with up to 63 sites
for the quantum chain, and ground state
considerations \cite{holtwes,holt} (checked by
Monte Carlo data) for the infinite classical chain. We considered
positive and negative single--ion anisotropies, $D$.

For $D > 0$, the supersolid or biconical phase is stable at
zero temperature.  As in the
case of $D/J= \Delta/2$, the broad BC phase
of the classical chain is effectively replaced, in the quantum chain,
by the corresponding, rather narrow SS phase as well as 
SL and (10) phases.

\begin{figure}[h]
\begin{minipage}{16pc}
\includegraphics[width=16pc]{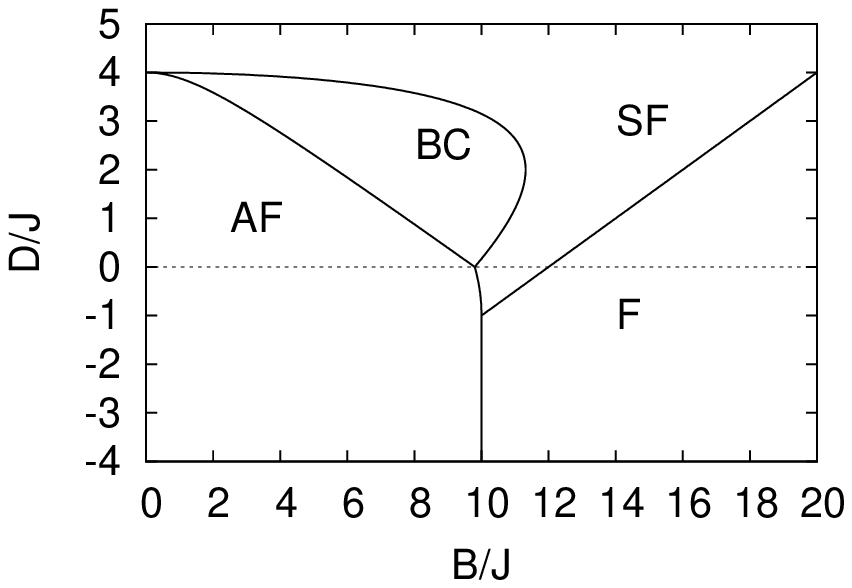}
\caption{\label{label} Ground state phase diagram of the classical
  infinite anisotropic Heisenberg chain with $\Delta= 5.0$.\\}
\end{minipage}\hspace{2pc}%
\begin{minipage}{16pc}
\includegraphics[width=16pc]{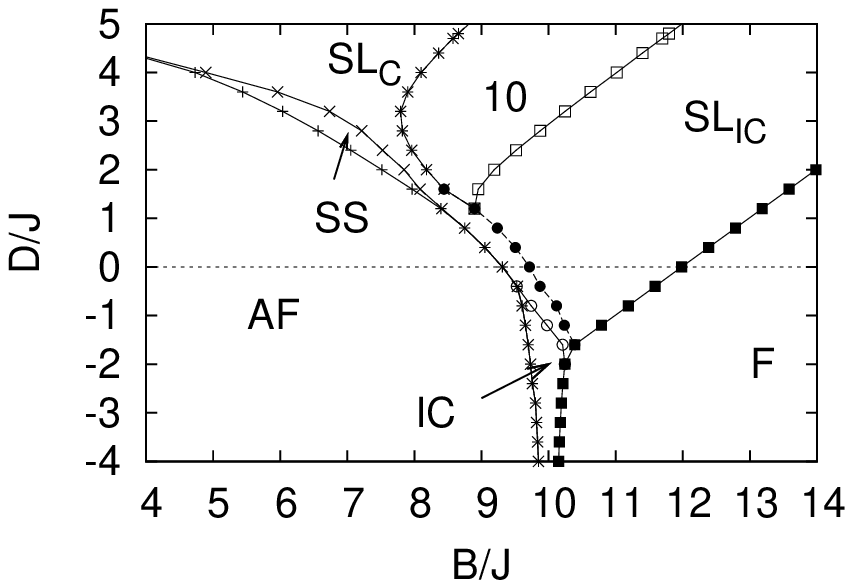}
\caption{\label{label}Ground state phase diagram of the spin--1
  anisotropic Heisenberg chain with $\Delta= 5.0$.\\}
\end{minipage} 
\end{figure}

In contrast to the case $D/J= \Delta/2$, the supersolid phase
is always bordered by the AF and
SL phases. Accordingly, we
observe only one type of magnetization profile. Illustrative
examples are depicted in Fig. 4, at $D/J= 3.0$ and various
fields. In the SS phase, at given single--ion anisotropy, $D$, and 
field, $B$, the magnetization takes on different values
at odd and even sites in the center of the
chain. The local magnetization $m_i$ tends to acquire the same
value at odd and even sites on approach to the SL phase in the quantum
chain. Actually, the classical BC phase  
is usually described by two tilt angles, with respect to the
$z$--axis for the two sublattices
formed by neighboring sites, with the tilt angles
approaching each other when getting closer to the SF
phase \cite{holtwes,holt}. Obviously, this behavior is
completely analogous to the one depicted in Fig. 4.

\begin{figure}[h]
\includegraphics[width=16pc]{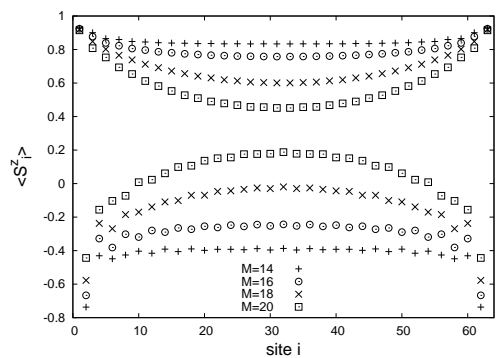}\hspace{2pc}%
\begin{minipage}[b]{16pc}\caption{\label{label}Magnetization profiles
in the supersolid phase at $\Delta= 5.0$, $D/J= 3.0$, and fields $B/J$
varying between 6.5 and 6.8. Spin chains with $L= 63$ sites
are considered. The lower symbols denote the even sites, the
upper ones the odd sites.}
\end{minipage}
\end{figure}

Increasing $D$, the 'large--D'
phase \cite{schulz,chen,mik,oitmaa} may eventually be
stable. It corresponds to the 
planar phase in the classical model with vanishing field, with the spin vectors
pointing perpendicular to the $z$--axis, being the ground state 
for $D/J \ge 4$. The new phase may be
expected to give rise to a SL phase at
non--zero fields. A discussion
of this interesting aspect is, however, beyond the scope of the present
contribution. 

For $D< 0$, the biconical or supersolid phase is no
longer stable. In fact, only AF, SL (or SF),
and F structures are encountered. The SL phase can be either
commensurate, with a wide plateau in the magnetization profile
away from the boundaries, or incommensurate, with modulations 
superimposed on the average magnetization. Obviously, there
are two IC phases, see Fig. 3. The one, denoted by 'IC' in
Fig. 3, occuring essentially in between 
the AF and F phases, has been found before, having
exponentially decaying transversal spin--spin
correlations \cite{sakai}, in contrast to the usual spin--liquids
with algebraic decay. It has no analog in the classical model, see
Fig. 2. The related transition between the IC and
SL$_C$  phases has been obtained before, being either of first
order or, at large average
magnetization, continuous \cite{sakai}. Our results agree with 
that description. We find another C--IC transition
line between the SL$_C$ and SL$_{IC}$ phases at somewhat larger fields
for given $D/J$, see Fig. 3 (full circles). Note that this
transition seems to take place at $M/L$ significantly larger
than 1/2 for $D <0$. Increasing $D$, $D> 0$, in the vicinity of
the (10) phase, the line goes over to the
above discussed scenario with commensurate, $M/L < 1/2$,  and
incommensurate, $M/L > 1/2$, structures. By further
increasing the 
planar single--ion anisotropy, $D/J$ being larger than
roughly 3.6, we observe in the 'SL$_C$' phase
close to the (10) phase modulated structures for rather 
short chains, $L \le 31$. The possible, additional C--IC border is not
displayed in Fig. 3. Indeed, a more detailed analysis, taking into
account finite--size effects, is desirable.  

In summary, we have studied ground state properties of
spin--1 antiferromagnetic anisotropic Heisenberg chains with
exchange and single--ion anisotropies in a field, for
given ratio of the two kinds of anisotropies, $D/J= \Delta/2$, and for
fixed exchange anisotropy, $\Delta= 5$, with varying single--ion
anisotropy. In both cases, supersolid phases are observed, in
accordance with the behavior of the corresponding 
classical spin chain, displaying biconical phases. The extent of
the supersolid phases is, however, substantially reduced as
compared to that of the biconical phases. Furthermore, the
spin--liquid phases show distinct commensurate and
incommensurate regions, presumably, separated by sharp
transitions. The corresponding spin--flop phases in
the classical model are always of commensurate type. In the
quantum model, a (10) phase appears as a ground state, being
present in the classical variant only in the Ising limit, the
Blume--Capel model.
  
\ack

We should like to thank C. D. Batista, M. Holtschneider, A. Kleine, A. Kolezhuk,
N. Laflorencie, P. Sengupta, J. Sirker, and S. Wessel for
useful discussions and information. The research has
been funded by the excellence initiative of
the German federal and state governments.

\end{document}